\documentclass[graybox]{svmult}

\usepackage{mathptmx}       % selects Times Roman as basic font
\usepackage{helvet}         % selects Helvetica as sans-serif font
\usepackage{courier}        % selects Courier as typewriter font
\usepackage{type1cm} 
\usepackage{amssymb}
\usepackage{amsmath}
\usepackage{mathrsfs}
\usepackage{makeidx}         % allows index generation
\usepackage{graphicx}        % standard LaTeX graphics tool
\usepackage{multicol}        % used for the two-column index
\usepackage[bottom]{footmisc}% places footnotes at page bottom

\makeatletter

\newcommand{\fmslash}[2][0mu]{%
  \mathchoice
    {\fmsl@sh\displaystyle{#1}{#2}}%
    {\fmsl@sh\textstyle{#1}{#2}}%
    {\fmsl@sh\scriptstyle{#1}{#2}}%
    {\fmsl@sh\scriptscriptstyle{#1}{#2}}}
\newcommand{\fmsl@sh}[3]{%
  \m@th\ooalign{$\hfil#1\mkern#2/\hfil$\crcr$#1#3$}}
\makeatother

\newcommand{\mptvec}{\not \!\! \vec{P}_T}

\makeindex             % used for the subject index
\begin{document}

\title*{Mass determination and event reconstruction at Large Hadron Collider}
\author{Abhaya Kumar Swain and Partha Konar}
\institute{Abhaya Kumar Swain \at Theoretical Physics Division, Physical Research Laboratory, Ahmedabad 380009, India, \email{abhaya@prl.res.in}
\and Partha Konar \at Theoretical Physics Division, Physical Research Laboratory, Ahmedabad 380009, India, \email{konar@prl.res.in}}
\maketitle

\abstract*{ After successful discovery of the Higgs boson, the Large Hadron Collider (LHC) would confront 
  the major challenge in searching for new physics and new particles. Any such observation necessitates 
  the determination of mass and other quantum numbers like spin, polarisation etc. Many of our theories 
  beyond the Standard Model (BSM) motivated from profound experimental indication of dark matter (DM), 
  trying to accommodate them as some stable BSM particles within these theory. In such scenario, 
  any production of heavy resonance of new particles eventually decay semi-invisibly resulting at least 
  two stable particles in the final state. Reconstruction of these events at hadron colliders together 
  with the mass determination of DM or intermediate particles is challenging and center to this present analysis. 
  In this work we discuss some mass restricting way that can lead us to determine the new particle mass 
  when it decays semi-invisibly. We will also present a new method which can be used for the full 
  reconstruction of the event in the above scenario.}

\abstract{ After successful discovery of the Higgs boson, the Large Hadron Collider (LHC) would confront 
  the major challenge in searching for new physics and new particles. Any such observation necessitates 
  the determination of mass and other quantum numbers like spin, polarisation etc. Many of our theories 
  beyond the Standard Model (BSM) motivated from profound experimental indication of dark matter (DM), 
  trying to accommodate them as some stable BSM particles within these theory. In such scenario, 
  any production of heavy resonance of new particles eventually decay semi-invisibly resulting at least 
  two stable particles in the final state. Reconstruction of these events at hadron colliders together 
  with the mass determination of DM or intermediate particles is challenging and center to this present analysis. 
  In this work we discuss some mass restricting way that can lead us to determine the new particle mass 
  when it decays semi-invisibly. We will also present a new method which can be used for the full 
  reconstruction of the event in the above scenario.}

\section{Introduction}\label{sec:1}
The remarkable discovery of Higgs boson by ATLAS~\cite{Aad:2012tfa} and CMS~\cite{Chatrchyan:2012ufa} 
at LHC has resolved a long standing question in the Standard Model (SM). Apart from acquiring more Higgs data 
to significantly improve the measurements, the prime goal of Large Hadron Collider (LHC) is to search for the
physics beyond Standard Model (BSM). Buoyed by significant number of experimental evidence on dark matter, most 
of the BSM theories accommodate dark matter (DM) as some new stable fundamental particle. 
Detection of these DM particles in LHC is challenging because it goes missing  in the detector 
and only manifest itself through missing energy signal. Additional complexity is introduced by the nature of 
hadron collider from the availability of partial informations of incoming parton momenta. 
Lot of efforts were gone to deal with missing transverse energy signals and to determine the quantities like mass, 
spin and polarisation of particles associated with those signals. For a recent review one can follow from 
ref.~\cite{Barr:2010zj, Barr:2011xt}. Among all the  inclusive and global mass determination 
variables $\hat{s}_{min}$ and its variants~\cite{Konar:2008ei, Konar:2010ma} are quiet interesting as they provide the 
mass scale of new physics without worrying about the production mechanism. These variables can also handle any number of 
invisibles and variety of them and simple analytical formula is also exists.

In this paper we try to formulate the usefulness of the topology information in the inclusive variable $\hat{s}_{min}$ 
by implementing the on shell constraints in the minimization. Principal consequence being the constrained phase space which
is bounded from below, as well as from above. So we define two new variables $\hat{s}_{min}^{cons}$ and $\hat{s}_{max}^{cons}$ 
which can be used as mass constraining variables. We also showed that $\hat{s}_{min}$ can be used for a unique event reconstruction 
for any topology and the momentum reconstruction can be improved significantly by these constrained variables. 
This paper is organized as follows: we will give a very brief introduction about the variable $\hat{s}_{min}$ 
and the constrained variables in section~\ref{sec:2}. We will discuss the event reconstruction capability of these variables in section~\ref{sec:3} followed by conclusion at section~\ref{sec:4}.

\section{$\hat{s}_{min}$ variable and its constrained counterpart using topology information}\label{sec:2}
 The partonic mandelstam variable $\hat{s}$ contains information like mass of heavy resonance for Antler topology or threshold of pair production of particles for non-Antler topology. Given a generic topology the system is under constrained, where it is nearly impossible to determine $\hat{s}$ experimentally which is the primary motivation to construct the variable $\hat{s}_{min}$ by minizing $\hat{s}$ w.r.t unknown invisible momenta subject to missing energy constraint. In other words $\hat{s}_{min}$ is the minimum partonic mandelstam variable that is consistent with all the visible particle momenta in the final state and missing transverse momentum constraints. It is a global and inclusive variable which makes it applicable to apply any topology with a simple analytical formula.
 
 \begin{figure}[t]
 \centering
 \includegraphics[scale=0.46,keepaspectratio=true]{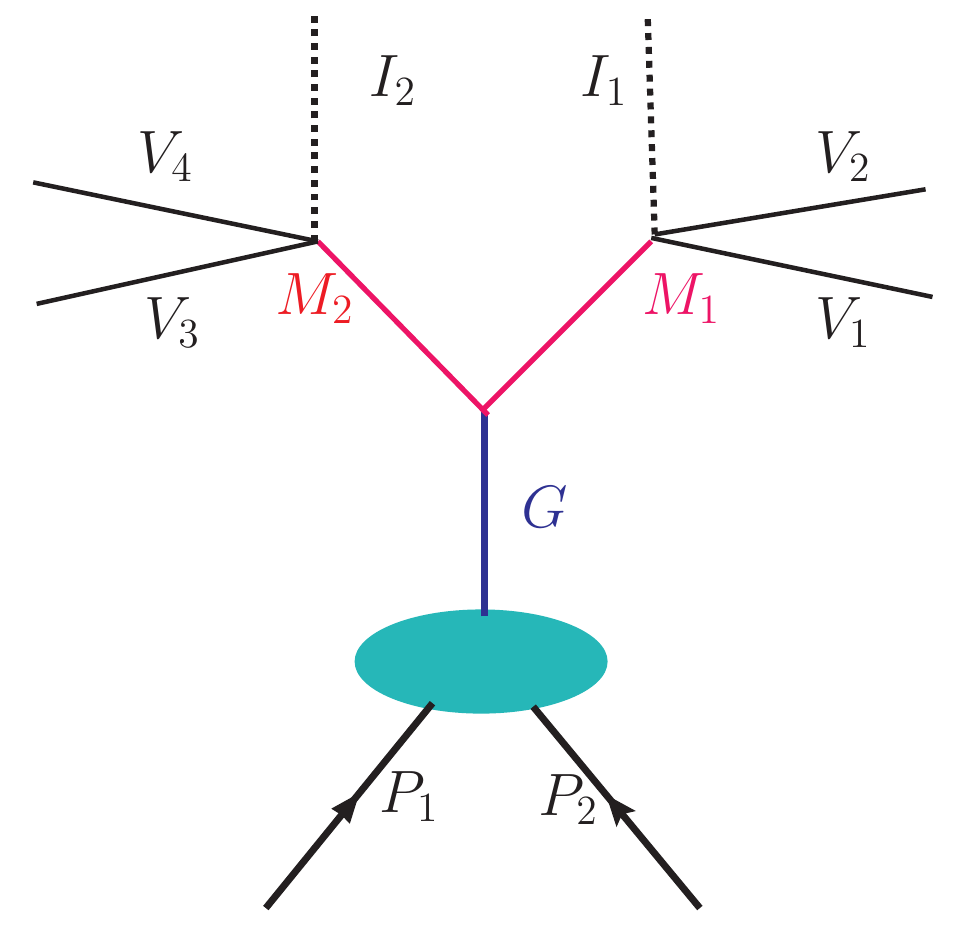}
 \includegraphics[scale=0.40,keepaspectratio=true]{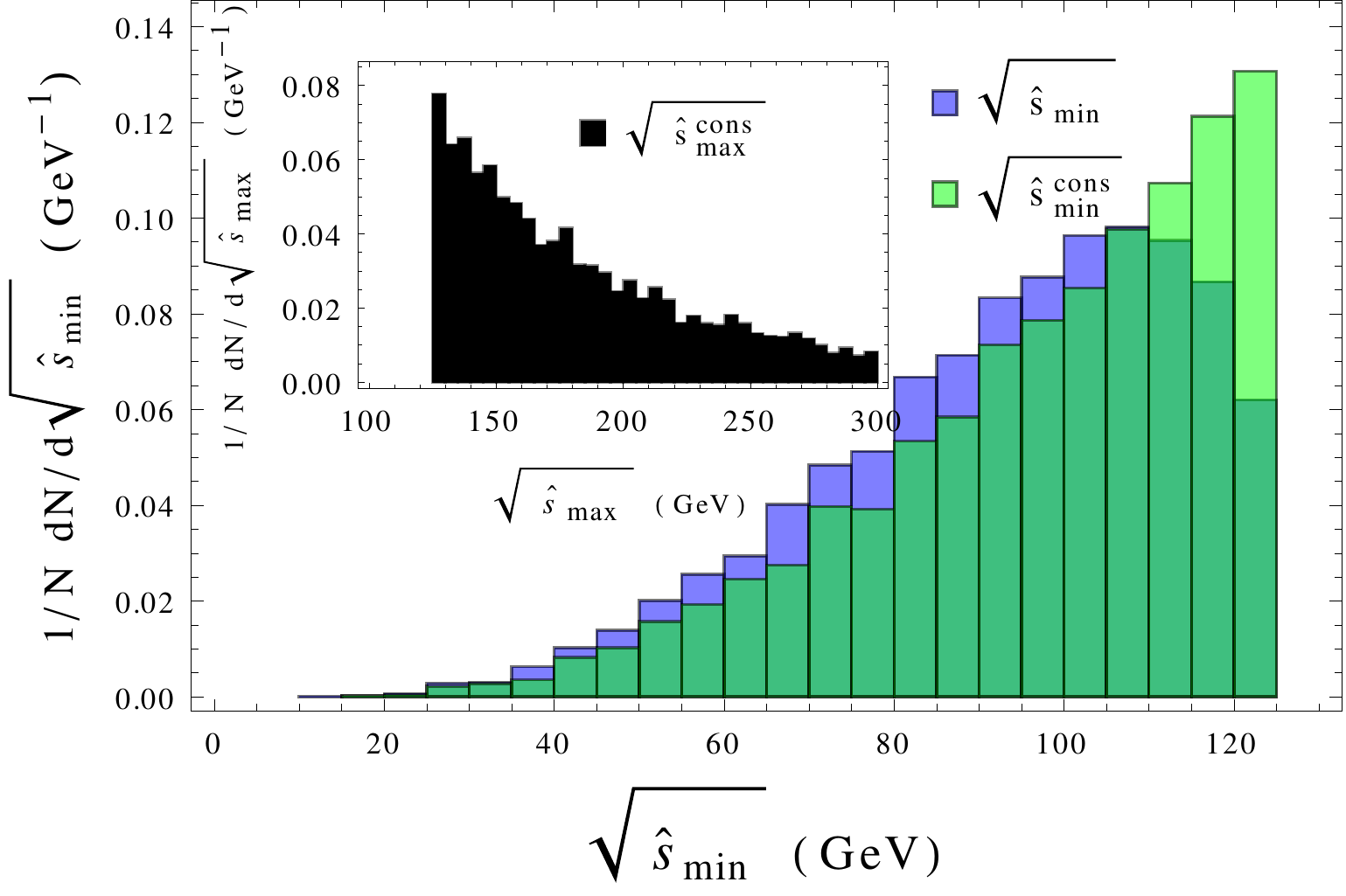}
 % Antler_topology.pdf: 471x352 pixel, 72dpi, 16.62x12.42 cm, bb=0 0 471 352
 \caption{The diagram on left represents Antler topology where G is a parity even resonance decays to two parity odd resonance $M_1$ and $M_2$ each of which further decays to two SM particles $V_k, k = 1, 2, 3, 4$ and one DM particle $I_i, i = 1, 2$. The right side figure shows distribution of $\hat{s}_{min}$ in blue, $\hat{s}_{min}^{cons}$ in green and the black figure inside it represents $\hat{s}_{max}^{cons}$.}
 \label{fig:antlertopology}
\end{figure}
 
 We simplify our discussion by making two assumptions which are common in BSM theories, that the dark matter stabilizing symmetry is a $Z_2$ symmetry and there is only one dark matter candidate in the theory under consideration. In this draft we restrict ourself for symmetric Antler topology as given in the figure~\ref{fig:antlertopology} where G is a $Z_2$ parity even heavy resonance decays to two parity odd resonance $M_1$ and $M_2$ each of which further decays to two SM particles $V_k$ and one DM particles $I_i$. The four momenta of visible and invisible particles are $p_k$ and $q_i$ respectively. The analytical formula for $\hat{s}_{min}$ and invisible momentum at the minimum for symmetric Antler topology can be found in the ref~\cite{Konar:2008ei}. The on shell constraints which $\hat{s}_{min}$ have not used in its minimization is the mass shell constraints of $M_1$ and $M_2$, we assume we know the masses\footnote{There are many examples in SM and beyond Standard Model theories as given in the ref.~\cite{Swain:2014dha} where we know the intermediate particles mass.} of these two resonance 
and use them to constrain the phase space. We referred to  $constraints$ as
 \begin{equation}
 constraints = \left \{
\begin{aligned}
   & (p_{1} + p_{2} + q_1)^2 = M_{M_1}^2, \,(p_{3} + p_{4} + q_2)^2 = M_{M_2}^2 \\ 
   & q_1^2 = M_{I_1}^2, \, q_2^2 = M_{I_2}^2 \\
   & \vec{q}_{1T} + \vec{q}_{2T} = \mptvec \label{missingptcons}.
   \end{aligned}
\right\}
 \end{equation}
 \{$M_{M_1}$, $M_{M_2}$\} and \{$M_{I_1}$, $M_{I_2}$\}  are the true masses of the intermediate particles \{$M_1$, $M_2$\} and the invisible particles \{$I_1$, $I_2$\} respectively. Now we define constrained variables using the equations in $constraints$ as
  \begin{eqnarray}
  \hat{s}_{min}^{cons} = \min_{\substack{\vec{q_1}, \vec{q_2} \\ \{constraints\}}} [\hat{s}(\vec{q_1}, \vec{q_2})]
 \end{eqnarray}
 \begin{eqnarray}
  \hat{s}_{max}^{cons} = \max_{\substack{\vec{q_1}, \vec{q_2} \\ \{constraints\}}} [\hat{s}(\vec{q_1}, \vec{q_2})]
 \end{eqnarray}
 $\hat{s}_{min}^{cons}$ is defined as the minimum of the constrained phase space which is bounded above by true of G. $\hat{s}_{min}^{cons}$ being the minimum of the constrained phase space it can not exceed the true mass of G, if it does that will point out that the constrained phase space does not contain the true mass point. $\hat{s}_{max}^{cons}$ defined as the  maximum of the constrained phase space which bounded below true mass of G because of the same reason given for $\hat{s}_{min}^{cons}$. The details about these variables can be found the ref.~\cite{Swain:2014dha}. The distribution of these variables with $\hat{s}_{min}$ is given in the figure \ref{fig:antlertopology}: blue histogram represents $\hat{s}_{min}$, green shows $\hat{s}_{min}^{cons}$ and the black histogram inside this figure is for  $\hat{s}_{max}^{cons}$.

\section{event reconstruction capability}\label{sec:3}

\begin{figure}[t]
 \centering
 \includegraphics[scale=0.35,keepaspectratio=true]{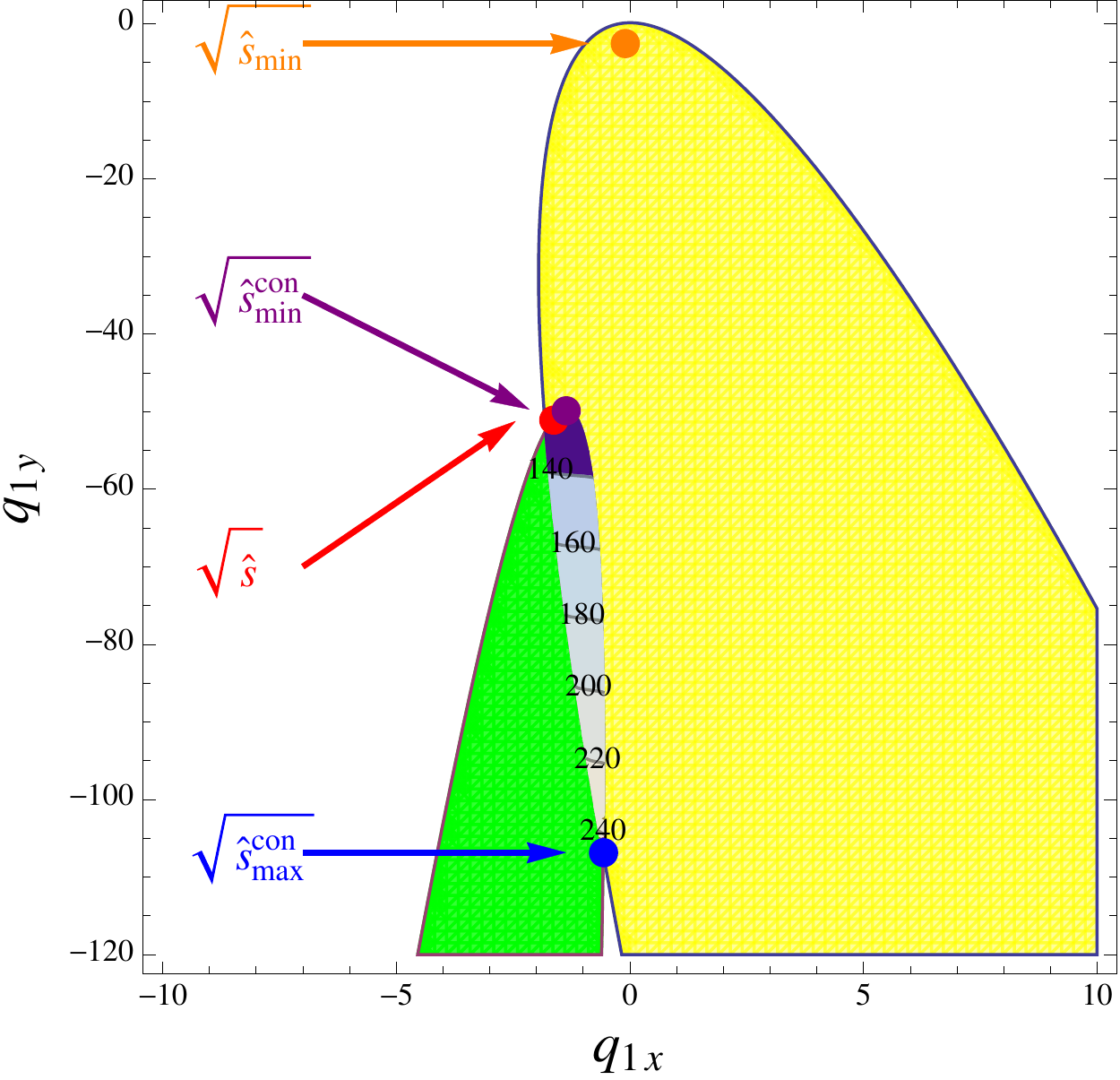}
 \includegraphics[scale=0.53,keepaspectratio=true]{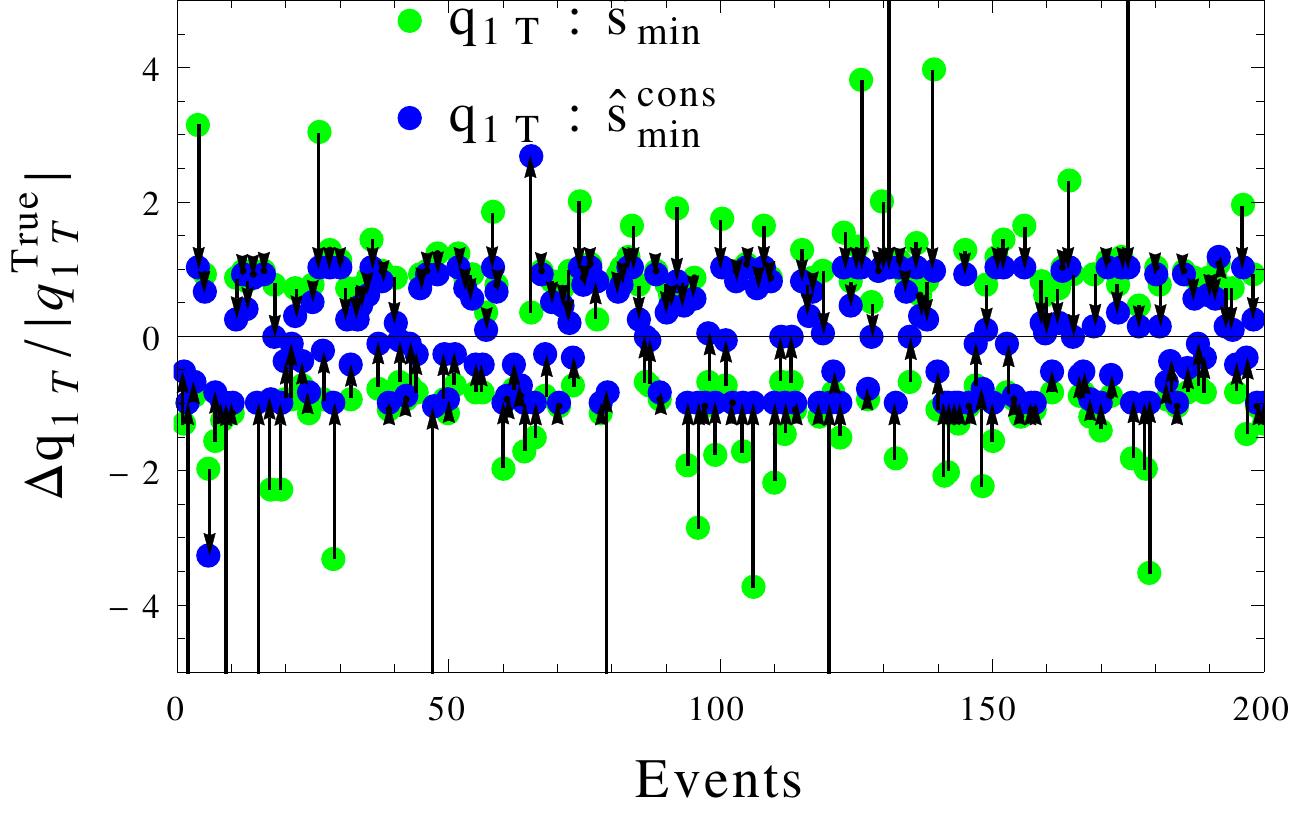}
 % constrained_shat_Antler_momentum_reco_shatConsMin.pdf: 360x347 pixel, 72dpi, 12.70x12.24 cm, bb=0 0 360 347
 \caption{Figure on left shows the constraint phase space coming from the intersection of the two ellipses  using the on shell constraints for one event. $\hat{s}_{min}$, $\hat{s}_{min}^{cons}$, $\hat{s}_{max}^{cons}$ and  $\hat{s}^{True}$ are presented by  orange, purple, blue and red points respectively. As can be seen the constrained variables performing better for this event. Figure on right shows for 200 events  the improvement of constrained variable over unconstrained one and also well correlated with true momenta.}
 \label{fig:q1shatconsmin}
\end{figure}

 In this section we describe the momentum reconstruction capability of $\hat{s}_{min}$  and its constrained counterpart $\hat{s}_{min}^{cons}$. One can also assign momenta to invisible particles using $\hat{s}_{max}^{cons}$, which can  be found in the ref.~\cite{Swain:2014dha}. In this paper we assign that momenta to invisible particles which gives $\hat{s}_{min}$ or $\hat{s}_{min}^{cons}$ that is  the  momenta from the minimization. Now  using the two assumptions described in section~\ref{sec:2} the momenta that gives $\hat{s}_{min}$ are, 
  \begin{eqnarray}
  q_{iT} &=&  \frac{1}{2}\mptvec,\\
  q_{iz} &=& \frac{1}{2} \frac{\mptvec}{\sqrt{(E^v)^2 - (P_z^v)^2}}\sqrt{M_{inv}^2 + \mptvec^2}.
 \end{eqnarray}
 Where $\mptvec$, $E^v$, $P_z^v$ and $M_{inv}$ are missing transverse momenta, total visible energy, total visible longitudinal momenta and sum of invisible particle masses respectively. We improve this momentum reconstruction by using on shell constraints of $B_1$ and $B_2$ in the minimization. The two on shell constraints of $B_1$ and $B_2$ are two ellipses and they are related by the missing energy constraints. So the intersection region satisfies all the six constraints (4 mass shell constraints + 2 missing energy constraints) and the minimum of that constrained phase space is called  $\hat{s}_{min}^{cons}$ and maximum is called $\hat{s}_{max}^{cons}$ which is shown in the left plot of Fig.~\ref{fig:q1shatconsmin}. In the right figure we have shown correlation between reconstructed momenta and  true momenta  for 200 events, where the green point shows the momenta from $\hat{s}_{min}$, blue point shows momenta from $\hat{s}_{min}^{cons}$ and black arrow points the motion of these points for each event.

\section{Conclusion}\label{sec:4}
\begin{itemize}
 \item $\hat{s}_{min}$ is a global and inclusive variable defined to measure  the mass scale of new physics and it only depends on the momenta of final state visible particles and missing transverse energy. The simple analytical formula of $\hat{s}_{min}$ allows one to calculate  the mass scale of new physics without worrying about the number of invisible and variety of them in the topology.
 
 \item In this paper we made two simplifying assumptions which are mostly common in BSM theories: The DM stabilization symmetry is a $Z_2$ symmetry resulting at least two invisible particle in the final state, there is only one DM candidate in the theory.  Now using these two assumptions one can approximate the unknown invisible momenta with the unique description from the minimization $\hat{s}_{min}$.
 %are unique and independent of trial mass of invisible particle but they are parallel which may not be the true configuration.
 
 \item We added the topology information and partial mass spectrum information in  $\hat{s}_{min}$ if we already have some of them. These additional input constraints the allowed phase space further. This system is such bounded, the minimum of constrained region is dubbed as $\hat{s}_{min}^{cons}$, bounded above by the true mass of G. Where as, the maximum of constrained region is dubbed as $\hat{s}_{max}^{cons}$, bounded below by the mass of G.  The relations between all such variables together with true mass are $\hat{s}_{min} \le \hat{s}_{min}^{cons} \le M_G \le \hat{s}_{max}^{cons}$.
 
 \item The constrained variables can also be used for event reconstruction and a comparison has been made between momentum reconstruction capability of constrained and unconstrained variables. Both $\hat{s}_{min}^{cons}$  significantly improves momentum reconstruction of invisibles giving better estimate for events near the endpoint, whereas, $\hat{s}_{max}^{cons}$ gives better momentum reconstruction for events near threshold.
\end{itemize}

%%%%%%%%%%%%%%%%%%%%%%%%%%%%%%%%%%%%%%%%%%%%%%%%%%% Bibliography %%%%%%%%%%%%%%%%%%%%%%%%%%%%%%%%%%%%%%%%%%%%%%%%

\end{document}